\documentclass[11pt,twoside]{article}
\usepackage{asp2010}

\resetcounters

\begin{document}
  \title{Binary Pulsar B1259-63 Spectrum Evolution 
      and Classification of Pulsar Spectra}
  \author{M. Dembska, J. Kijak, W. Lewandowski
  \affil{Kepler Institute of Astronomy, University of Zielona G\'ora, Lubuska 2, 65-265 Zielona G\'ora, Poland}}

 \begin{abstract}
  Recently published results \citep{kijak2011b} indicated
the evidence for a new aspect in radio pulsars spectra. We studied the
radio spectrum of PSR B1259-63 in an unique binary with Be star LS 2883
and showed that this pulsar undergoes a spectrum evolution due to orbital motion. 
We proposed a qualitative model which explains this evolution. We considered
two mechanisms that might influence the observed radio emission: free-free
absorption and cyclotron resonance. According to published results \citep{kijak2011a},
 there were found objects with a new type of pulsar radio spectra,
called gigahertz-peaked spectra (GPS) pulsars. Most of them were found
to exist in very interesting environments. Therefore, it is suggested that the
turnover phenomenon is associated with the environment than being related
intrinsically to the radio emission mechanism. Having noticed the apparent
resemblance between the B1259-63 spectrum and the GPS, we suggested that
the same mechanisms should be responsible for both cases. Thus, we believe
that this binary system can hold the clue to the understanding of gigahertz-peaked
spectra of isolated pulsars. Using the same database we constructed spectra for chosen
observing days and obtained different types of spectra. Comparing to current
classification of pulsar spectra, there occurs a suggestion that the appearance
of various spectra shapes, different from a simple power law which is typical for
radio pulsars, is possibly caused by environmental conditions around neutron
stars.
 \end{abstract}

  \section{The Gigahertz-peaked Spectra Pulsars}
\begin{figure}[!ht]
 \centering
  \includegraphics[width=0.55\textwidth]{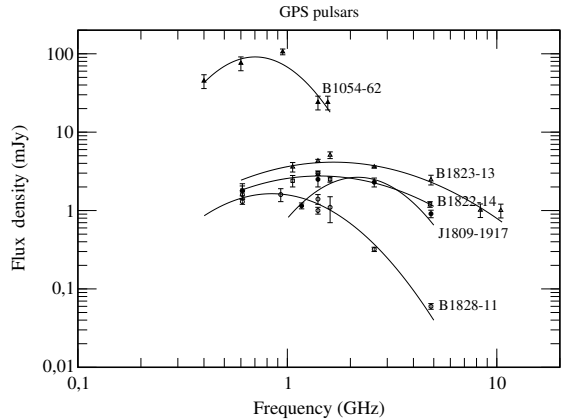}
  \caption{The spectra of five known GPS pulsar (Kijak et al. 2007 and 2011).}
  \label{fig1}
\end{figure}

Generally, the observed radio spectra of most pulsars can be modelled as a power law with negative spectral indices of about -1.8 \citep{maron2000}. If a pulsar can be observed at frequencies low enough (i.e. \mbox{100-600 MHz}), it may also show a low-frequency turnover in its spectrum \citep{sieber1973,malofeev1994}. On the other hand, \citet{lorimer1995} mentioned three pulsars
which have positive spectral indices in the frequency range 300-1600 MHz. Later, \citet{maron2000} re-examined spectra of these pulsars taking into account the data obtained at higher frequencies (above 1.6 GHz). 
In paper \citet{kijak2007}, the authors presented the first direct evidence for turnover in pulsar radio spectra  at high frequencies. Based on their observations of these pulsars, 
\citet{kijak2011a} provided a definite evidence for a new type of pulsar radio spectra. These spectra show the maximum flux above 1 GHz, while at higher frequencies the spectra look like a typical pulsar spectrum.  At lower frequencies (below 1 GHz), the observed flux decreases, showing a positive spectral index (see. Fig. \ref{fig1}). A frequency at which such a spectrum shows the maximum flux was called the peak frequency.
They called these objects the gigahertz-peaked spectra (GPS) pulsars. \citet{kijak2011a} also indicated that the GPS pulsars are  relatively young objects, and they usually adjoin such interesting environments as HII regions or compact pulsar wind nebulae.  Additionally, some of they seem to be coincident with the known but sometimes unidentified X-ray sources from third EGRET Catalogue or HESS observations. We can assume that the GPS appearance owes to the environmental conditions around the neutron stars rather than to the radio emission mechanism.

\section{PSR B1259-63 Spectrum Evolution}
PSR B1259-63 was also inscribed by \citet{lorimer1995} in the list of pulsars with positive spectral indices. Therefore,  seems a natural candidate to be classified as the GPS pulsar. 
\begin{figure}[!ht]
 \centering
 \begin{tabular}{cc}
   \includegraphics[width=0.45\textwidth]{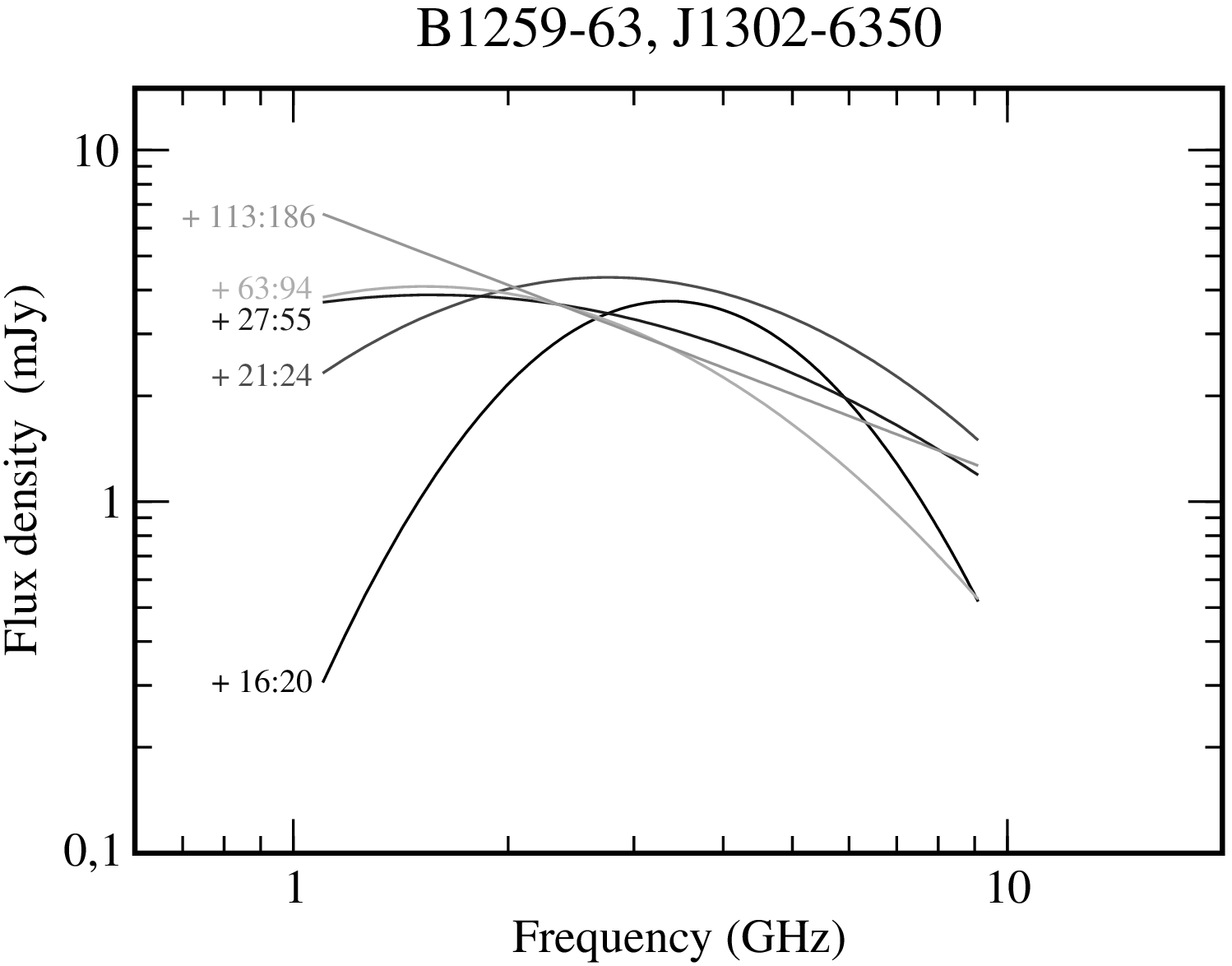}
   \includegraphics[width=0.45\textwidth]{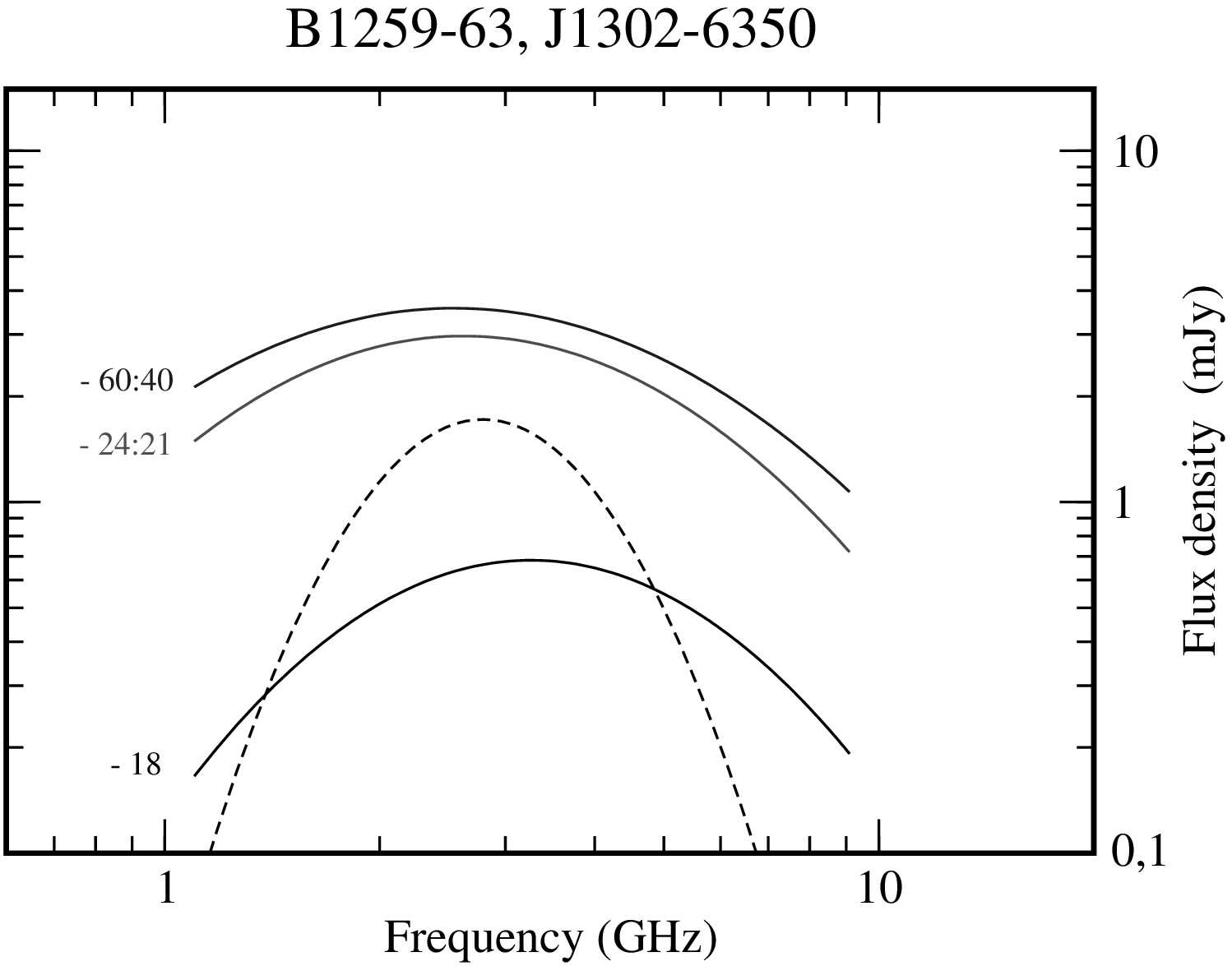}\\
 \end{tabular}
 \caption{The fits to the B1259-63 spectra for the each orbital phase range,
from 60 d prior to periastron (left panel) up to 186 d after it (right panel). }
 \label{fig2}
\end{figure}

PSR B1259-63 has a short period of 48 ms and a characteristic age of 330 kyr. Its average dispersion measure (DM)  is about 147 pc cm$\mathrm{^{-3}}$ and the corresponding distance is about 2.75 kpc. 
The companion star LS 2883 is a 10-mag massive Be star with a mass of about 10M$\mathrm{_{\odot}}$  and a radius of 6R$\mathrm{_{\odot}}$.  
Be stars are generally believed to have a hot tenuous polar wind and a cooler high-density equatorial disc.
We studied the radio spectrum of B1259-63 and showed that the shape of the spectrum depends on the orbital 
phase \citep{kijak2011b}. We analysed the available measurements of the pulsed flux obtained during three periastron  passages (1997, 2000 and 2004) which allowed us to study shapes of the pulsar 
spectrum in details (see Fig. \ref{fig2}). Our analysis showed that this pulsar undergoes a spectrum evolution 
due to orbital motion. We suggested that this effect is caused by the interaction of the radio waves with the Be star environment. In addition, we have shown that the peak frequency also depends on the orbital phase and therefore 
it varies with the changes of the pulsar environment. We argued that such behaviour can be explained by the radio-wave absorption in the magnetic field associated with the disc. We proposed a qualitative model which explains this evolution. We argued that the observed variation of the spectra is caused by a combination of two  effects: the free-free absorption in the stellar wind and the cyclotron resonance in the magnetic field. 
This field is associated with the disc and is infused by the relativistic particles of the pulsar wind.
\section{B1259-63 Spectrum: Detailed Study and Classification of Pulsar Spectra}
Using the same database \citep{johnston1999, connors2002, johnston2005} we constructed spectra for chosen observing days. 
\begin{figure}[!ht]
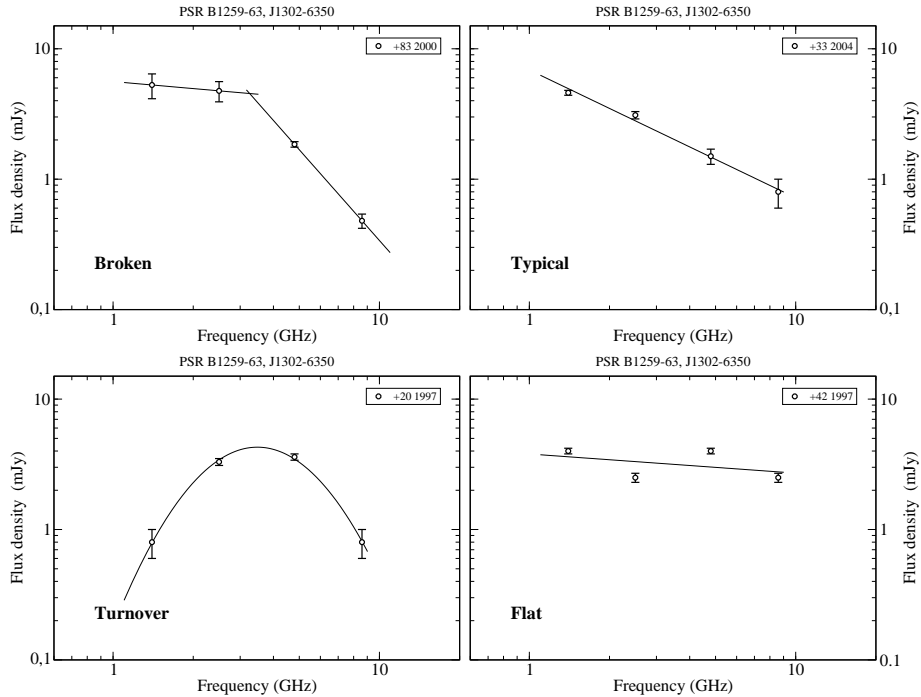

 \centering
 \begin{tabular}{cc}
   \includegraphics[width=0.45\textwidth]{b1259-63_broken.eps}
   \includegraphics[width=0.45\textwidth]{b1259-63_typical.eps}\\
   \includegraphics[width=0.45\textwidth]{b1259-63_turnover.eps}
   \includegraphics[width=0.45\textwidth]{b1259-63_flat.eps}\\
 \end{tabular}
 \caption{The fits to the B1259-63 spectra for chosen days. Each panel shows different type of spectra.}
 \label{fig3}
\end{figure}
As previously \citep{kijak2011b} we analysed the shapes of the B1259-63 spectrum at various orbital phase ranges and obtained results are consistent with those for intervals.
The flux at the given frequency apparently changes with orbital phases.
When the pulsar is close to periastron, the flux generally decreases at all observed
frequencies and the most drastic decrease is observed at the lowest frequency.
\section{Conclusions}
Having noticed the apparent resemblance between the B1259-63 spectrum and the GPS, we suggested that the same mechanisms should be  responsible for both cases \citep{kijak2011b}. The only difference could be an invariable shape of the GPS, in contrast to the B1259-63  spectrum, which undergoes evolution due to orbital motion. We can concluded that the GPS feature should be caused by some external factors  rather than by the emission mechanism. 
Thus, the case of B1259-63 can be treated as a key factor to explain the GPS phenomenon observed for the solitary pulsars with interesting environments.

Close to the periastron point the spectra of B1259-63 resemble those of the GPS pulsars. 
The spectrum for the orbital epochs further from the periastron point are more consistent 
with typical pulsar spectra (i.e. power-law and broken). Moreover, detailed study of PSR B1259-63 spectra
revealed the appearance of all types of spectral shapes, including a flat
spectrum (see Fig. 3).

We believe that the case of B1259-63 can be treated as a key factor to our
understanding of not only the GPS phenomenon (observed for the solitary
pulsars with interesting environments) but also other types of untypical
spectra as well (e.g. flat or broken spectra). This in turn would suggest,
that the appearance of various non-standard spectra shapes in the general
population of pulsars can be caused by peculiar environmental conditions.

\acknowledgements On of the authors (MD) is a scholar within Sub-measure 8.2.2 Regional Innovation Strategies, Measure 8.2 Transfer of knowledge, Priority VIII Regional human resources for the economy Human Capital Operational Programme co-financed by European Social Fund and state budget.
\bibliography{b1259-63_ERPM_proceedings}

 \end{document}